# Reflections and Thoughts on Tired Light

M. Moore & J. Dunning-Davies,
Department of Physics,
University of Hull,
Hull HU6 7RX,
England.

email: **j.dunning-davies@hull.ac.uk**

**Abstract.**

The position of the various tired light theories is reviewed briefly and it is noted that one of the biggest objections to them concerns the mechanism by which light might lose energy as it travels through space. Here some new work relating to the constancy of the speed of light is highlighted as providing a possible solution to this conundrum, thus making more feasible explanation of phenomena via theories involving the notion of tired light.

# Introduction.

As Disney said not too long ago [1]:
> '*Cosmology rests on a very small database: it suffers from many fundamental difficulties as a science (if it is a science at all) whilst observations of distant phenomena are difficult to make and harder to interpret. It is suggested that cosmological inferences should be tentatively made and sceptically received.*'

These sentiments may not meet with the approval of many, but a moment's reflection should convince most open-minded people that they contain at least an element of truth. Nowadays, it sometimes appears that the Big Bang model for the origin of the Universe is accepted as established fact, rather than simply another theory – albeit one with a multitude of ardent supporters and which seems to explain so much so satisfactorily. However, problems do remain and many have been addressed in the past by allowing additions to the basic theory – a privilege not afforded to rival theories. Amongst the rival theories is the Steady State theory espoused by Hoyle, Gold and Bondi. This theory has been upgraded to allow for new knowledge and the results have been written up and presented to the scientific community for consideration [2] but it certainly doesn't seem to have received a very open hearing even though, at the very least, it raises once again some very interesting points such as who originally detected the cosmic background radiation and which theories do, or do not, explain this phenomenon. Various other theories have been advanced in attempts to explain some or all of the problems of cosmology and one, which possibly deserves further contemplation due to advances in knowledge, must be the so-called tired light theory. An added reason for not dismissing this out of hand is the intellect of some of those advocating it as a serious theory; - people such as Walther Nernst and Max Born for example. This is a point brought out quite forcibly in the article by Assis and Neves [3], an article to which reference might be made profitably.

The phenomenon of the redshift plays an important role in cosmology and astrophysics. It is a well-known and thought to be well-understood effect in which the wavelength of electromagnetic radiation is lengthened as a result either of the source moving away from the observer – the so-called Doppler effect – or by the actual expansion of the Universe. The effect due to movement is well documented but the usual expressions have to be modified for high speeds to take account of special relativistic effects. However, the redshift caused by the expansion of the Universe has nothing to do with the Doppler effect but is thought to be caused by the expansion of space itself, which is felt to stretch the wavelength of the radiation that is travelling towards the observer. Normally, of course, the radiation under consideration is light. Frequently, it seems to be accepted that this gravitational redshift is something predicted by the General Theory of Relativity but, as has been shown [4], there is no need to introduce that theory to explain this effect. The idea of tired light, the notion that light somehow loses energy in its journey through the cosmos, came into being as an alternative explanation for models which included an expanding Universe and suggested that, as the light lost energy, it became redshifted. One big success the idea enjoyed was in predicting a temperature of $2.8^{o}K$ for the cosmic background radiation at a time when the Big Bang models predicted temperatures anywhere between $5^{o}K$ and $50^{o}K$. Some, however, felt the theory not to be an extension of known physics but rather an ad hoc addition which, although fitting some known facts, offered no reasonable explanation for them. Some wondered, not unreasonably, why light should lose energy as it travels through space. Hence, tired light theories have tended to be forgotten, but is this sensible and should they be re-examined?

**Tired Light Theories.**

As early as 1912, Nernst had proposed the notion of the Universe being in a steady state. By 1937 he had developed the idea further and had suggested an explanation for the cosmological redshift in terms of tired light; that is, Nernst was suggesting that the æther absorbed radiation, thus causing a decrease in the energy and, therefore, the frequency of galactic light. In other words, Nernst was ruling out the Doppler effect explanation for this observed phenomenon of redshift. It is important to note the use of the Stefan-Boltzmann law, which is characteristic of black body radiation, in both Nernst's reasoning and that of others at that time in scientific history. For example, Eddington crucially relied on the Stefan-Boltzmann law in his derivation of a value for the temperature of space. Later, in the 1950's, a tired light model was proposed by Finlay-Freundlich to explain the redshift of solar lines and anomalous redshifts of some stars as well as the cosmological redshift. This work was examined further by Max Born who suggested that the new effect could be due to a photon-photon interaction. One major problem with this suggestion is that it is not in agreement with presently accepted theory. However, it is still an interesting view to consider.

Most textbooks today regard the Big Bang theory as offering the true explanation for the origin of the Universe. The existence of the Steady State theory is usually mentioned but dismissed. Often the important piece of evidence supporting this is the prediction by Gamow and his collaborators of the $2.7^o K$ temperature of the cosmic background radiation before its discovery by Penzias and Wilson, while the rival Steady State theory did not predict this temperature. As is shown quite clearly in the book by Hoyle, Burbidge and Narliker [2], none of the statements in the previous sentence is true in fact and, while those authors might be felt to hold a vested interest in such a claim, it is one supported by other writers, such as Assis and Torres [3], and by the independent evidence. Both these models do, however, accept the Doppler shift interpretation of the cosmological redshift and so, both accept the idea of the expansion of the Universe. The third possibility of a Universe in dynamical equilibrium with neither expansion nor continuous creation of matter as proposed and developed by such as Nernst, Finlay-Freundlich and Born does exist still, however, and should not be dismissed completely out of hand. One outstanding problem with the tired light theories is, though, the identification of the physical process which brings about this energy loss for the quanta of light. The search for such a mechanism continues and Pecker and Vigier came up with a possibility in 1988 [5]. They drew attention to the possibility of photons interacting with vacuum particles resulting in a loss of energy for the photons. They also drew attention to a very important attitude of mind in asserting that such 'exotic' theories should be viewed with open minds. They also noted that the popular Big Bang theory has had many additions made to it and, as such, has lost much of the simplicity that was probably the greatest asset of the original.

The question of a mechanism for tired light is, however, still an open one amongst those who don't summarily dismiss the theory. Hence, the area is one which should attract some attention.

## Suggestions for Consideration.

It must be recognised that one obvious objection to Nernst's ideas will be his reference to an æther. However, recently, the notion of an æther has been resurrected in an attempt to offer an explanation for the apparent 'missing matter' in our Universe [6]. Actually, the idea of a luminiferous æther has never truly left science but it is only very recently that it has begun to be acceptable to talk of it again. In his 1985 article [7], Thornhill showed that the æther is an ideal gas and that the intrinsic energy of a mass $m$ of æther is $9mc^2/4$. This is, incidentally, a figure which links very well with estimates of the magnitude of the dark energy necessary to support the Big Bang. This indicates that minds should not remain totally closed to ideas and that possibly the notion of an æther may be useful in explaining some observed phenomena. Crucially in the present context, if the existence of an æther is not completely dismissed, Nernst's ideas concerning the origin of the redshift must be reconsidered and this should be done with open minds unhampered by preconceived conclusions.

One apparently major problem encountered when considering questions about light is the popularly held belief that Einstein's special theory of relativity precludes any variation in the speed of light. It is often stated and widely believed that one of the bases of relativity is the constancy of the speed of light. *However, Einstein's assumption concerning the speed of light actually referred to the speed of light in a vacuum. In any case, it is well known experimentally that the speed of light simply isn't constant, but actually varies according to the medium* through which it is passing. This experimental knowledge immediately raises the question of exactly how the speed of light varies. Again in his 1985 article, Thornhill [7] showed that it must vary with the square root of the background temperature, which immediately implies that it varies in time and would, in fact, slow down with the passage of time. However, even if this is accepted, it does not explain why it should have different values depending on the medium through which it is passing and certainly does not preclude it depending on some other variable as well. Accordingly, Santilli [8] has suggested that it may be a function of the refractive index of the material through which it is passing. The effects of this suggestion are quite far-reaching and offer possible explanations for a number of phenomena which have been providing food for thought for astrophysicists for some considerable time.

Santilli notes that the famous equation
$$E = mc^2$$
is strictly valid only for point particles moving in a vacuum – conditions enunciated quite clearly by Einstein himself. Hence, the above relation is not necessarily valid for extended, deformable, non-spherical particles. According to Santilli's new isorelativity theory, under these circumstances the above relation should be replaced by
$$\widehat{E} = mC^2 = mc^2/n^2,$$
where $C$ represents the image of the speed $c$ within the interior of the medium under consideration and where $n$ is less than one. This immediately suggests two alternative approaches. In one the energy might be assumed to remain the same in the generalisation from special relativity to isorelativity. If this assumption is made, the mass and speed of light must change. Alternatively, the mass might be thought to remain unchanged; in which case, it is the energy and speed of light which must change. Since the value of $C$ is likely to be much greater than $c$ inside truly dense media, the energy equivalence of a given mass in isorelativity could be much greater than the usual Einsteinian value.

For light, it is usual to consider energy, $E$, related to frequency, $\nu$, via
$$E = h\nu.$$
Combining the above equations indicates immediately how the frequency of the light may be related to the speed and, if this speed varies with the refractive index of the medium through which that light is passing, it follows that the frequency, and hence the wavelength, will be related to that refractive index also. The light moving towards us from some distant source will be emitted from a body of greater density than the medium through which it passes subsequently. Hence, according to this theory, its speed will decrease as it passes through what is euphemistically referred to as space, resulting in an increase of wavelength. In other words, light moving towards us from some distant source will not move with constant speed and its wavelength will be altered on its journey towards the earth, resulting in an observed redshift.

The ideas of Santilli in this and other fields are relatively new and knowledge of them is not as widespread as possibly it should be, but this is a field of thought that should not be dismissed out of hand. Rather it should be considered seriously and with an open mind, always remembering that, using these ideas, Mignani [9] has been able to offer a perfectly feasible explanation for Arp's observations of quasars with high redshifts which appeared physically linked with galaxies apparently possessing much lower redshift values. At the same time, the wording of Einstein's original assumption should be viewed carefully and, in future, possibly more care should be taken when quoting the assumption regarding the constancy of the speed of light.

## References.